\documentclass[twocolumn,epsfig,superscriptaddress,showpacs,oneside,amsmath,amssymb,floatfix]{revtex4}

\usepackage{epsfig,multirow}

\begin{document}

\title{Orbital-transverse density-wave instabilities in iron-based
  superconductors}

\author{Zi-Jian Yao}
\affiliation{Department of Physics and Center of Theoretical and Computational Physics, The University of Hong Kong, Pokfulam Road, Hong Kong, China}

\author{Jian-Xin Li}
\affiliation{National Laboratory of Solid State Microstructures and Department of Physics, Nanjing University, Nanjing 210093, China}

\author{Q. Han}
\affiliation{Department of Physics, Renmin University of China, Beijing, China}
\affiliation{Department of Physics and Center of Theoretical and Computational Physics, The University of Hong Kong, Pokfulam Road, Hong Kong, China}

\author{Z. D. Wang}
\affiliation{Department of Physics and Center of Theoretical and Computational Physics, The University of Hong Kong, Pokfulam Road, Hong Kong, China}

\date{\today}
\begin{abstract}

  Besides the conventional spin-density-wave (SDW) state, a new kind
  of orbital-transverse density-wave (OTDW) state is shown to exist
  generally in multi-orbital systems. We demonstrate that the orbital
  character of Fermi surface nesting plays an important role in
  density responses. The relationship between antiferromagnetism and
  structural phase transition in LaFeAsO (1111) and BaFe$_2$As$_2$
  (122) compounds of iron-based superconductors may be understood in
  terms of the interplay between the SDW and OTDW with a five-orbital
  Hamiltonian. We propose that the essential difference between 1111
  and 122 compounds is crucially determined by the presence of the
  two-dimensional $d_{xy}$-like Fermi surface around (0,0) being only
  in 1111 parent compounds.

\end{abstract}


\maketitle

\section{introduction}
Over the last two years, research on iron-based superconductors has
been an exciting topic that attracts intensely
experimental~\cite{kamihara_08,chen_08,ren_08,chen:247002,rotter:107006}
and
theoretical~\cite{han:37007,daghofer:237004,chen:047006,cvetkovic:024512,chubukov:134512,yu:064517}
investigations. Although it has been well established experimentally
that this family of compounds exhibit several phase transitions
including the structural phase transition
~\cite{zhao:953,mcguire:094517,krellner:100504}, the
antiferromagnetic (AF) phase
transition~\cite{dong:27006,mcguire:094517,krellner:100504}, and the
superconducting phase transition, the mechanism of these phase
transitions remains highly controversial. However, there has been a
consensus on the basic Fermi surface (FS) topology - hole pockets
centered at $(0,0)$ and electron pockets centered at
$(\pi,\pi)$~\cite{ding:47001,sato:047002,liu:177005}. From a viewpoint of
itinerant antiferromagnetism, the hole and electron pockets are
assumed to be nested nearly perfectly in the parent compound. The
spin-density wave (SDW) state is stabilized due to the existence of
the on-site Coulomb repulsions. Upon doping, the long range AF order
is destroyed and short range AF spin fluctuations are developed, which
is responsible for the high-temperature
superconductivity~\cite{yao:025009,wang:047005}.

The multi-orbital nature of iron-based superconductors is believed to
play a prominent role in the superconductivity, which signifies the
importance of extracting the distinct physics that emerges from
multi-orbital effects in this new family of materials. In this paper,
we reveal that a new kind of unconventional orbital-transverse
density-wave state exists generally in multi-orbital systems with
certain orbital configuration of FS nesting. This type of density wave
stems from the rotation asymmetry of Hamiltonian in the orbital space,
which is reflected in the multi-orbital FS nesting geometry, and has
an intriguing impact on the electron charge and spin density
responses. In connection to the iron-based superconductors, our
calculations show that the orbital-transverse density wave is a
competing order with conventional SDW.

This paper is organized as follows. In Section~\ref{dw}, we first
consider three prototypes of FS nesting in the multi-orbital systems.
Then we define the density operators with two orbital degrees of
freedom, and show how their responses are affected by the orbital
configuration of FS nesting. In Section~\ref{iron}, we show that the
orbital-transverse density wave is nearly degenerated with the SDW by
using a realistic five-orbital Hamiltonian of the iron-based
superconductors. The interplay between the structure distortion and
antiferromagnetism is discussed in the context of competing
density-wave ground states. Finally, several remarks are drawn as the
summary in Section~\ref{sum}.

\section{density waves with orbital degrees of freedom}
\label{dw}
\subsection{Multi-orbital Hamiltonian}
A multi-orbital Hamiltonian incorporating the on-site intra- and
inter-orbital Coulomb interaction reads
\begin{equation}
  H=H_0+\frac{1}{2}U\sum_{i\alpha\sigma}n_{i\alpha\sigma}n_{i\alpha\bar{\sigma}}+\frac{1}{2}U^{\prime}\sum_{i,\alpha{\neq}\beta}n_{i\alpha}n_{i\beta},
  \label{ham1}
\end{equation}
where $
H_0=\sum_{i{\alpha}j{\beta}\sigma}t_{i{\alpha},j{\beta}}c_{i\alpha\sigma}^{\dagger}c_{j\beta\sigma}
$, $t_{i{\alpha},j{\beta}}$ is the hopping term between orbital
$\alpha$ of site $i$ and orbital $\beta$ of site $j$, and
$n_{i\alpha\sigma}$ the electron number operator, $U$ and $U^{\prime}$
is the on-site intra- and inter-orbital Coulomb repulsion,
respectively. For simplicity, $U$ is assumed to be equal to
$U^{\prime}$ in the present study and the terms of Hund's rule
coupling are ignored. In order to illustrate various types of density
responses of the multi-orbital system more clearly, we start from a
simple two-orbital model in the square lattice with a hole pocket
lying at $(0,0)$ and an electron pocket lying at $(\pi,0)$. The
dispersion and chemical potential ensure that the electron pocket and
hole pocket are circular and nested perfectly. This fermiology is
compatible with many possible orbital configurations, which have
different orbital-weights of the FS. In Fig. 1 we show three
prototypes of them. As will see below, the density response behaviors
of them are quite different.


\begin{table*}
  \renewcommand{\arraystretch}{1.3}
  \caption{Density waves with two orbital degrees of freedom}
  \centering \begin{tabular}{c|c|rl|c|c}
    \hline {\bf Ordering type} & {\bf Orientation in} & \multicolumn{2}{|c|}{\multirow{2}{*}{\bf Density operator}} & \multirow{2}{*}{\bf Response function} & {\bf Divergent} \\ {\bf of densities} & {\bf orbital space} & & & & {\bf blocks} \\ \hline \hline
     total spin& n/a & $D_{0,3}(r_i)$: & $\rho_{i1\uparrow}-\rho_{i1\downarrow}+\rho_{i2\uparrow}-\rho_{i2\downarrow}$
    & $\frac{1}{2}(\chi^s_{1111}+\chi^s_{2222}+\chi^s_{1122}+\chi^s_{2211})$ &  $\hat{\chi}^{s1}$ \\ \hline
     total charge& n/a & $D_{0,0}(r_i)$: & $\rho_{i1\uparrow}+\rho_{i1\downarrow}+\rho_{i2\uparrow}+\rho_{i2\downarrow}$ & $\frac{1}{2}(\chi^c_{1111}+\chi^c_{2222}+\chi^c_{1122}+\chi^c_{2211})$ & $\hat{\chi}^{c1}$ \\ \hline
    orbital-
    & longitudinal & $D_{3,3}(r_i)$: & $\rho_{i1\uparrow}-\rho_{i1\downarrow}-(\rho_{i2\uparrow}-\rho_{i2\downarrow})$ & $\frac{1}{2}(\chi^s_{1111}+\chi^s_{2222}-\chi^s_{1122}-\chi^s_{2211})$ & $\hat{\chi}^{s1}$ \\
    polarized spin & transverse & $D_{-(+),3}(r_i)$: & $c^{\dagger}_{i2\uparrow}c_{i1\uparrow}-c^{\dagger}_{i2\downarrow}c_{i1\downarrow}$ (H.c.) & $\chi^s_{2121}$ & $\hat{\chi}^{s2}$ \\ \hline
    orbital- & longitudinal & $D_{3,0}(r_i)$: & $\rho_{i1\uparrow}+\rho_{i1\downarrow}-(\rho_{i2\uparrow}+\rho_{i2\downarrow})$ & $\frac{1}{2}(\chi^c_{1111}+\chi^c_{2222}-\chi^c_{1122}-\chi^c_{2211})$ & $\hat{\chi}^{c1}$ \\
    polarized charge & transverse & $D_{-(+),0}(r_i)$: & $c^{\dagger}_{i2\uparrow}c_{i1\uparrow}+c^{\dagger}_{i2\downarrow}c_{i1\downarrow}$ (H.c.) & $\chi^c_{2121}$ & $\hat{\chi}^{c2}$ \\ \hline
  \end{tabular}
  \label{table1}
\end{table*}

\subsection{Definition of density quantities}
The introduction of orbital degrees of freedom enables us to define
extra physical quantities of density whose translational symmetry
could be broken when the system is phase transferred into the
density-wave ground state. Simiar to the spin-$\frac{1}{2}$ degrees of
freedom, we represent the two-orbital degrees of freedom by an
$\frac{1}{2}$ pseudospin. The density operators in such a spin-orbital
space can be written as
$\mathbf{D}(r_i)=\mathbf{\phi}_i^{\dagger}\mathbf{\Gamma}\mathbf{\phi}_i$ with
$\mathbf{\Gamma}=\mathbf{\tau}\otimes\mathbf{\sigma}$ and
$\mathbf{\phi}_i^{\dagger}=[{c_{i1\uparrow}^{\dagger},c_{i1\downarrow}^{\dagger},c_{i2\uparrow}^{\dagger},c_{i2\downarrow}^{\dagger}}]$
the 4-component spinor ($\tau^0=\sigma^0=\mathbb{I}$ and
$\tau^{1-3}=\sigma^{1-3}$ the Pauli matrices, where $\tau$ is defined
in the orbital space and $\sigma$ is defined in the spin space). To
study the instability in the particle-hole channels, the sixteen
particle-hole operators
$\phi^{\dagger}(\tau^{0-3}\otimes\sigma^{0-3})\phi$ can be reduced
into six different density channels (three of spin and three of
charge) due to the SU(2) symmetry of Hamiltonian~\eqref{ham1} in the
spin space, which are listed in Table~\ref{table1}. 

The six channels listed in Table~\ref{table1} could be further grouped
into four types of density waves, if we are not interesed in the
direction of the density orbital-polarization, namely the total
spin-density wave, the total charge-density wave, the
orbital-polarized spin-density wave, and the orbital-polarized
charge-density wave. Basically the orbital-polarized density operator
represents the difference of density between two orthogonal orbitals.
The longitudinal or transverse polarization of orbital determines
which two orthogonal orbitals are density-ordered, for instance, the
longitudinal polarization corresponds to a density-ordered state
between orbital 1 and 2 while the transverse polarization corresponds
to a density-ordered state between two orthogonal mixtures of orbitals
1 and 2. The distinction between orbital-longitudinal and
orbital-transverse parts of the susceptibilities is consequent to the
rotation asyemmetry of the Hamltonian~\eqref{ham1} in the orbital
space. In Fig. 2 we plot schematically the ground states of these
density-wave states in real space.

\begin{figure}
  \epsfig{figure=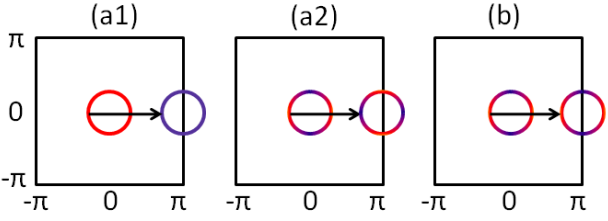,width=0.45\textwidth}
  \caption{(Color online) Schematic plot of the Fermi surfaces with different orbital
    configurations. Blue and red indicate dominant weights of orbital
    1 and 2, respectively.}
\end{figure}

\begin{figure}
  \epsfig{figure=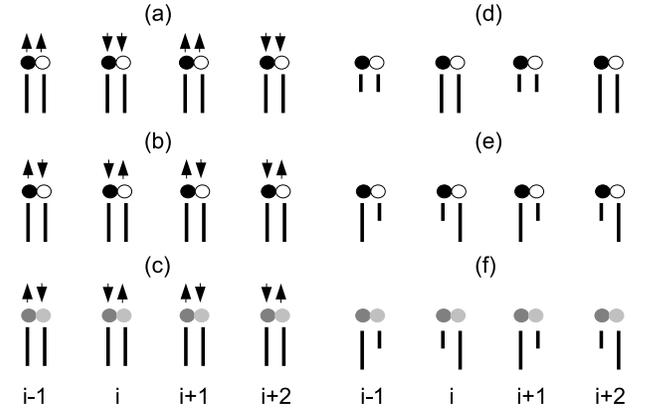,width=0.45\textwidth}
  \caption{Schematic plot of density waves with wave vector
    $Q=(\pi,0)$, (a)-(f) correspond to the density-wave states with
    density operators $D_{0,3}$, $D_{3,3}$, $D_{-(+),3}$, $D_{0,0}$,
    $D_{3,0}$, and $D_{-(+),0}$. Black (blank) circle represents
    orbital 1 (2). Dark grey and grey circles represent two orthogonal
    mixtures of orbitals 1 and 2, e.g.,
    $\frac{1}{\sqrt{2}}(\left|1\right\rangle+\left|2\right\rangle)$
    for dark grey and
    $\frac{1}{\sqrt{2}}(\left|1\right\rangle-\left|2\right\rangle)$
    for grey. The arrow indicates spin density and the vertical line
    indicates the charge density. $i$, $i+1$, ... denote the lattice
    sites and each site represents a reduced unit cell. Only modulated
    direction is shown.}
\end{figure}
\label{definition}

\subsection{Instability criterions}
The particle-hole instabilities in the presence of Coulomb
interactions may be examined within the random-phase approximation
(RPA). The interaction matrices for the magnetic ($\hat{U}^s$) and
charge ($\hat{U}^c$) channels reads
\begin{equation}
  \label{eq:tew}
\hat{U}^s=\left(\begin{array}{cc} \hat{U}^{s1} & 0\\
0 & \hat{U}^{s2}\end{array}\right),
\hat{U}^c=\left(\begin{array}{cc} \hat{U}^{c1} & 0\\
0 & \hat{U}^{c2} \end{array}\right),
\end{equation}
where $\hat{U}^{s1}=$ diag$(U,U)$, $\hat{U}^{c1}_{mn}=U$ for $m=n$ and
$2U^{\prime}$ otherwise,
$\hat{U}^{s2}=$ diag$(U^{\prime},U^{\prime})=-\hat{U}^{c2}$. Defining
the orbital indices of the bare susceptibility as
$\chi_{\alpha\beta\mu\nu}^0(q)=-\frac{T}{N}\sum_{k}G^0_{\alpha\mu}(k+q)G^0_{\nu\beta}(k)$,
we get the RPA spin and charge susceptibility matrices
\begin{eqnarray}&&
\label{chirpa}
\hat{\chi}^s({q})=[\hat{I}-\hat{\chi^0}(q)\hat{U}^s]^{-1}\hat{\chi^0}(q), \nonumber \\&&
\hat{\chi}^c({q})=[\hat{I}+\hat{\chi^0}(q)\hat{U}^c]^{-1}\hat{\chi^0}(q).
\end{eqnarray}
The response functions listed in Table~\ref{table1} are given by
\begin{eqnarray}
  \tilde{\chi}_{A,B}(q,i\omega)&=&\frac{1}{2}\int_{0}^{\beta}d{\tau}e^{i\omega\tau}{\langle}T_{\tau}D_{A,B}(q,\tau)D_{A,B}(-q,0)\rangle \nonumber \\
  \tilde{\chi}_{-+,B}(q,i\omega)&=&\frac{1}{2}\int_{0}^{\beta}d{\tau}e^{i\omega\tau}{\langle}T_{\tau}D_{-,B}(q,\tau)D_{+,B}(-q,0)\rangle \nonumber \\
  \label{chi_s}
\end{eqnarray}
where $A$ and $B$ are numbers and
$D_{A,B}=\phi^{\dagger}(\tau_{A}\otimes\sigma_{B})\phi$. A direct
calculation gives the response functions of density operators defined
in Table~\ref{table1}
\begin{eqnarray}
  \tilde{\chi}_{0,3}&=&\frac{1}{2}(\chi^s_{1111}+\chi^s_{2222}+\chi^s_{1122}+\chi^s_{2211}),\nonumber \\
  \tilde{\chi}_{0,0}&=&\frac{1}{2}(\chi^c_{1111}+\chi^c_{2222}+\chi^c_{1122}+\chi^c_{2211}),\nonumber \\
  \tilde{\chi}_{3,3}&=&\frac{1}{2}(\chi^s_{1111}+\chi^s_{2222}-\chi^s_{1122}-\chi^s_{2211}), \nonumber \\
  \tilde{\chi}_{-+,3}&=&\chi^s_{2121},\nonumber \\
  \tilde{\chi}_{3,0}&=&\frac{1}{2}(\chi^c_{1111}+\chi^c_{2222}-\chi^c_{1122}-\chi^c_{2211}), \nonumber \\
  \tilde{\chi}_{-+,0}&=&\chi^c_{2121}.
\end{eqnarray}

To a good approximation, especially at $Q=(\pi,0)$, the bare
susceptibility matrix in the two-orbital case can be regarded as a
block matrix $\hat{\chi}^0=$
diag$(\hat{\chi}^1,\hat{\chi}^2)$, where
\begin{eqnarray}
\hat{\chi}^1=\left(\begin{array}{cc} \chi^0_{1111} & \chi^0_{1122}\\
    \chi^0_{2211} & \chi^0_{2222}\end{array}\right),
\hat{\chi}^2=\left(\begin{array}{cc} \chi^0_{1212} & \chi^0_{1221}\\
    \chi^0_{2112} & \chi^0_{2121}\end{array}\right) \nonumber.
\end{eqnarray}
As a result, the RPA spin and charge susceptibility matrices are also
blocked, 
\begin{equation}
\hat{\chi}^{s(c)}=\left(\begin{array}{cc} \hat{\chi}^{s1(c1)} & 0\\
0 & \hat{\chi}^{s2(c2)}\end{array}\right),
\end{equation}
where
\begin{equation}
\label{chirpa}
\hat{\chi}^{s1}=\frac{\hat{M}_1}{d_1}, \hat{\chi}^{s2}=\frac{\hat{M}_2}{d_2},
\hat{\chi}^{c1}=\frac{\hat{M}_3}{d_3}, \hat{\chi}^{c2}=\frac{\hat{M}_4}{d_4}.
\end{equation}
Here $M_n$ are some $2\times2$ matrices and
\begin{small}
\begin{eqnarray}
\label{denom}
d_1&=&{\rm det}[\hat{\chi}^{s1}]=(1-U\chi^0_{1111})(1-U\chi^0_{2222})-U^2\chi^0_{1122}\chi^0_{2211}, \nonumber \\
d_2&=&{\rm det}[\hat{\chi}^{s2}]=(1-U^{\prime}\chi^0_{1212})(1-U^{\prime}\chi^0_{2121})-U^{\prime2}\chi^0_{1221}\chi^0_{2112}, \nonumber \\
d_3&=&{\rm det}[\hat{\chi}^{c1}]=1+U(\chi^0_{1111}+\chi^0_{2222})+2U^{\prime}(\chi^0_{1122}+2\chi^0_{2211}) \nonumber \\
&&+(U^2-4U^{\prime2})(\chi^0_{1111}\chi^0_{2222}-\chi^0_{1122}\chi^0_{2211}) \nonumber \\
d_4&=&d_2
\end{eqnarray}
\end{small}
are the denominators that come from the inversion of matrices
$\hat{I}-\hat{\chi^0}(q)\hat{U}^s$ and
$\hat{I}+\hat{\chi^0}(q)\hat{U}^c$ in Eq.~\eqref{chirpa}. Deviated
from the single band RPA, there are two instability criterions for the
spin channel, which are indicated by $d_1\rightarrow0$ and
$d_2\rightarrow0$, respectively. When approaching the critical
temperature, either the upper block $\chi^{s1}$ or the lower block
$\chi^{s2}$ may diverge. Comparing the response functions listed in
Table~\ref{table1} and Eq.~\eqref{chirpa}, we can see that the
divergence of the upper block or the lower block corresponds to a
transition to one or the other totally different magnetic phase, i.e.,
the former indicates an SDW transition while the latter indicates an
orbital-transverse polarized spin-density-wave transition. The same
conclusion holds for the charge susceptibility matrix. We will refer
the orbital-transverse density-wave ground state as OTDW in this
paper hereafter.

Now we turn to the real-space distribution of these density waves. As
discussed in Section~\ref{definition}, it is clear that the
orbital-longitudinal operators represent the density differences
between orbitals 1 and 2. On the other hand, the real-space
distribution of the orbital-transverse density wave can be revealed by
performing a rotation in the orbital space:
\begin{equation}
  \psi_i^{\dagger}=[{c_{i+\uparrow}^{\dagger},c_{i+\downarrow}^{\dagger},c_{i-\uparrow}^{\dagger},c_{i-\downarrow}^{\dagger}}], \nonumber
\end{equation}
where 
\begin{equation}
  c^{\dagger}_{i{\pm}\sigma}=\frac{1}{\sqrt{2}}(c^{\dagger}_{i1\sigma}{\pm}c^{\dagger}_{i2\sigma}). \nonumber
\end{equation}
The charge and spin density differences between the two orbitals
$\left|+\right\rangle$ and $\left|-\right\rangle$ can be represented
by
\begin{eqnarray}
  D^{\prime}_{3,3}(r_i)&=&\psi_i^{\dagger}(\tau_{3}\otimes\sigma_{3})\psi=S_{i+,z}-S_{i-,z}\nonumber \\
  &=&c_{i1\uparrow}^{\dagger}c_{i2\uparrow}-c_{i1\downarrow}^{\dagger}c_{i2\downarrow}+{\rm H.c.}, \nonumber \\
  D^{\prime}_{3,0}(r_i)&=&\psi_i^{\dagger}(\tau_{3}\otimes\sigma_{0})\psi=\rho_{i+}-\rho_{i-}\nonumber \\
  &=&c_{i1\uparrow}^{\dagger}c_{i2\uparrow}+c_{i1\downarrow}^{\dagger}c_{i2\downarrow}+{\rm H.c.}, \nonumber
\end{eqnarray}
where $S_{i\pm,z}$ and $\rho_{i\pm}$ represent the spin and charge
density of orbital $\left|{\pm}\right\rangle$ at $i$ site,
respectively. Similarly, the response functions for density operators
$D^{\prime}_{A,B}$ are
\begin{eqnarray}
  \tilde{\chi}^{\prime}_{3,3}&=&\frac{1}{2}(\chi^s_{1212}+\chi^s_{2121}+\chi^s_{1221}+\chi^s_{2112}),\nonumber \\
  \tilde{\chi}^{\prime}_{3,0}&=&\frac{1}{2}(\chi^c_{1212}+\chi^c_{2121}+\chi^c_{1221}+\chi^c_{2112}).
\label{trans_res}
\end{eqnarray}
Therefore, in OTDW state, the translational symmetry of the local
density difference between the two orthogonal mixtures of orbitals 1
and 2 is broken.

\subsection{Nesting-induced OTDW: a weak-coupling analysis}
Roughly speaking, the density response of a Hubbard-like Hamiltonian
is mainly determined by two factors: one is the band structure, and
the other is the Coulomb interactions. We now look into the influence
of different band structures shown in Fig. 1 on the density responses.
The bare susceptibility of a general multi-orbital system is give by
\begin{eqnarray}
  \chi^0_{\alpha{\beta}\mu\nu}(q,0)=&&\frac{1}{N}\sum_{k,mn}a^{\alpha*}_{m}(k+q)a^{\beta}_{n}(k)
  a^{\nu*}_{n}(k)a^\mu_{m}(k+q)\nonumber \\ &&\frac{f[\epsilon_{n}(k+q)]-f[\epsilon_{m}(k)]}{\epsilon_{m}(k)-\epsilon_{n}(k+q)+i\eta},
  \label{chi0}
\end{eqnarray}
where $\alpha$, $\beta$, $\mu$ and $\nu$ are the orbital indices, $m$
and $n$ are the band indices, and $a$ is the orbital weight. For the
case indicated in Fig. 1 (a1), the matrix elements of bare
susceptibility are simplified to
\begin{equation}
\chi^0_{\mu\nu\mu\nu}(q,0)=\frac{1}{N}\sum_{k,\mu\nu}\frac{f[\epsilon_{\nu}(k+q)]-f[\epsilon_{\mu}(k)]}{\epsilon_{\mu}(k)-\epsilon_{\nu}(k+q)+i\eta}.
\end{equation}
The susceptibilities $\chi^0_{\mu\mu\nu\nu}$ ($\mu\neq\nu$) vanish
since there is no hybridization between the two orbitals.
Straightforwardly, $\chi^0_{1212}(Q)$ and $\chi^0_{2121}(Q)$ are
divergent due to FS nesting, while $\chi^0_{1111}$ and $\chi^0_{2222}$
keep finite as they take no advantage from the nesting between
electron and hole pockets. Bear in mind that although the cases of
Fig. 1 (a1) and (a2) have different orbital configurations, their FS
nestings are both from the inter-orbital contribution~\cite{hybri}.
Hence the above conclusion holds also for the band structure shown in
(a2). For the case Fig. 1 (b), from Eq.~\eqref{chi0} the divergent
components of bare susceptibilities are $\chi^0_{1111}$ and
$\chi^0_{2222}$.

The above considerations focus on the band structure, but do not take
into account the electron-electron interactions explicitly. Given the
interaction vertices in Eq.~\eqref{eq:tew} and the instability
criterions in Eq.~\eqref{denom}, three spin-density waves ($D_{0,3}$,
$D_{3,3}$, and $D_{-(+),3}$), and one charge-density wave
($D_{-(+),0}$) could be established in Hamiltonian~\eqref{ham1}. The
$D_{0,0}$ and $D_{3,0}$ charge-density waves are prohibited due to
repulsive on-site Coulomb interactions (note that $d_3$ is always
greater than zero when $U\chi^0_{1111/2222}<1$ and the hybridization
is neglectable). $D_{0,3}$ and $D_{3,3}$ density waves occur
simultaneously since they both come from the divergent of the upper
block of $\hat{\chi}^{s}(Q)$, which indicates the real-space
distribution of spin density to be compatible with Fig. 2 (a) and (b).
A typical pattern of spin density that is either $D_{0,3}$ or
$D_{3,3}$ density wave is shown in Fig. 3, where the staggered spin
density is orbital-polarized and dominantly appears in one orbital. On
the other hand, the phase transition in $D_{-(+),0(3)}$ channel comes
from the divergent of the lower block of $\hat{\chi}^{s(c)}(Q)$ which
is coupled to the inter-orbital repulsion, with order parameter $
\left\langle
  c^{\dagger}_{\mu\sigma}(k+Q)c_{\nu\bar{\sigma}(\sigma)}(k)
\right\rangle$. In the present model, instabilities in $D_{-(+),3}$
and $D_{-(+),0}$ channels are degenerated due to $U=U^{\prime}$, while
the actually established state depends on the additional interaction
terms that break the degeneracy, e.g., the former is favored by Hund's
rule coupling while the latter is favored by the presence of
electron-phonon interactions.

We now conclude that for configurations shown in Fig. 1 (a1) and (a2),
the divergent susceptibility matrix is $\chi^{s2}(Q)$ or
$\chi^{c2}(Q)$, indicating that the system undergos an OTDW
transition. There is no long range order of the total local spin
density, as is shown in Fig. 2 (c) and (f). For configuration Fig. 1
(b), the divergent susceptibility matrix is $\chi^{s1}(Q)$,
corresponding to a spin density-wave transition whose real-space
distribution is shown in Fig. 3.

\begin{figure}
  \epsfig{figure=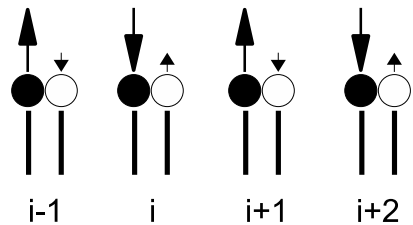,width=0.3\textwidth}
  \caption{Intra-orbital nesting induced spin-density-wave state that
    is realized in the iron-based superconductors, where the
    translational symmetries of $\left\langle{D}_{0,3}\right\rangle$
    and $\left\langle{D}_{3,3}\right\rangle$ are broken
    simultaneously. All the symbols are the same as that in Fig. 2.}
\end{figure}

\section{OTDW and the iron-based superconductors}
\label{iron}
The multi-orbital band structure of iron-based superconductors
provides an intermediate example between prototypes of Fig. 1 (a) and
(b). To be more concrete, we here employ a realistic five-orbital
Hamiltonian with on-site intra- and inter-orbital Coulomb repulsions.
The involved orbitals are $d_{xz}$, $d_{yz}$, $d_{x^2-y^2}$, $d_{xy}$,
and $d_{3z^2-r^2}$ (defined in the reduced unit cell) which are
labeled with 1, 2, 3, 4, and 5 respectively. Then the orbital indices
in Hamiltonian~\eqref{ham1} run from 1 to 5. We adopt the hopping
parameters given by Graser {\it et al}~\cite{graser:025016} fitted
from first principle calculated band structure by Cao {\it et
  al}~\cite{cao:220506}. The density-wave instabilities defined in
Table~\ref{table1} can be realized in the space spanned by the two most
relevant orbitals. In our calculations, $d_{xz}/d_{yz}$ (orbital 1/2)
and $d_{xy}$ (orbital 4) are the two orbitals with largest intra- and
inter-orbital susceptibilities. The instabilities of OTDW and SDW are
found to be nearly degenerated, consequently the competition between
them is sensitive to the detailed band structure. We consider this to
be a key point to understand the essential difference between {\it
  R}FeAsO (1111) and {\it A}Fe$_2$As$_2$ (122) compounds ({\it
  R}$=$rare earth and {\it A}$=$Sr, Ca, Ba and K), which will be
discussed in the following paragraphs.

The AF and structural transitions in the iron-based superconductors
exhibit different properties in different systems: the static AF order
develops after the structure distortion in 1111
compounds~\cite{mcguire:094517}, while in the 122 compounds, the two
transitions occur at the same
temperature~\cite{krellner:100504,zhao:140504}. The structure
distortion has been theoretically proposed to be driven
magnetically~\cite{yildirim:057010,fang:224509,xu:020501} or
electrically~\cite{turner:3782,lv:224506}. Here we present an
alternative view. Note that all the density waves we discuss are
nesting-driven and occur at the same wave-vector $Q=(\pi,0)$. The
order parameters of these density waves and the
tetragonal-to-orthorhombic distortion break the same symmetry.
Therefore, we propose that either spin-polarized SDW (shown in Fig. 3)
or OTDW could induce a structure distortion. When the transition
temperature of SDW $T_{d1}$ is higher than that of OTDW $T_{d2}$, the
AF transition is orbital-polarized as shown in Fig. 3 and is
accompanied with a structure distortion which is magnetically driven,
and the critical temperature of structure distortion $T_S$ and AF
transition $T_{AF}$ equal to $T_{d1}$. The temperature gap between
$T_S$ and $T_{AF}$ appears when $T_{d2}>T_{d1}$, which indicates
$T_S=T_{d2}>T_{AF}=T_{d1}$. We make two remarks here. First,
$D_{-(+),0}$ and $D_{-(+),3}$ density waves are two candidates for
OTDW-induced lattice distortion. As an orbital-ordered state of
charge, $D_{-(+),0}$ density wave could induce a structure distortion
electrically by Jahn-Teller effect. The coupling between $D_{-(+),0}$
density wave and lattice distortion is, if exists, magnetic relevant
and more subtle. It is out of scope of the present work to determine
which of the two density waves is {\it actually} established in the
real materials of iron-based superconductors. Since the two states are
degenerated in our model, we just generally related the structure
distortion with OTDWs. Second, in this scenario, from high temperature
to low temperature, the system undergos two phase transitions, both
couple to a structure distortion. However, only one structure
distortion is identified in experimental measurements. We note the
lattice distortion is weak and driven by breaking of $C_4$ lattice
rotation symmetry. Thus if the system is already in the orthorhombic
structure phase, another density-wave phase transition with
wave-vector $(\pi,0)$ would not break an extra lattice symmetry.

\begin{figure}
  \epsfig{figure=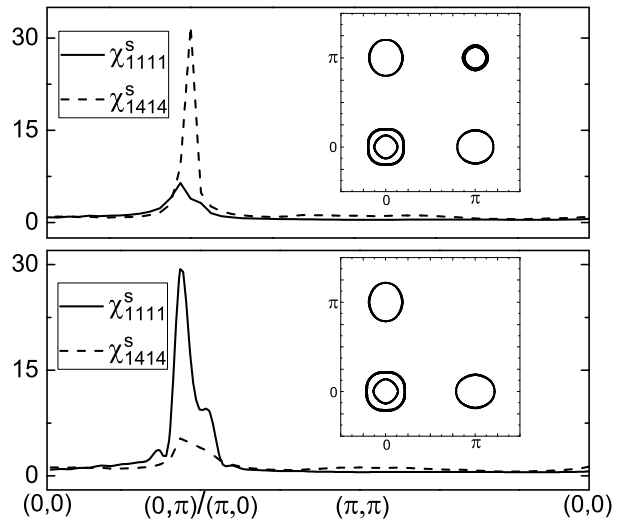,width=0.45\textwidth}
  \caption{Dominant components of spin susceptibility matrix with
    (upper) and without (lower) the appearance of $\varepsilon$-FS.
    The hopping parameters are adopted from~\cite{graser:025016},
    $U=U^{\prime}=1.38$ and $T=0.19$ for the upper panel and
    $U=U^{\prime}=1.53$ and $T=0.17$ for the lower panel. To produce
    the $\varepsilon$-FS, the on-site energy of orbital $d_{xy}$ is
    adjusted from 0.3 to 0.38 and $\mu$ is adjusted to 0.006 in the
    upper panel. All energy quantities are in units of eV.}
\end{figure}

It is known from band calculations of 1111 compounds that along
$\Gamma$-$Z$ direction lies a $d_{xy}$-like two-dimensional (2D) FS
(denoted as $\varepsilon$-FS hereafter)~\cite{vildosola:064518}. While
for 122 compounds, a 2D FS with $d_{xy}$ character is found to be
absent when the pnictogen height is relaxed using total-energy
minimization~\cite{singh:094511}. In our model calculation we find the
existence of $\varepsilon$-FS enhances the inter-orbital
susceptibility $\chi^0_{1414}$ substantially, while without
$\varepsilon$-FS, the susceptibility $\chi^0_{1111}$ is in general, at
least sightly, stronger than that of $\chi^0_{1414}$. The enhancement
of $\chi^0_{1414}$ by the appearance of a $d_{xy}$-like FS is expected
because of the strengthened inter-orbital nesting between
$\varepsilon$-FS and the $d_{xz}/d_{yz}$-dominant FS sections around
M. In Fig. 4 we show dominant components (1111 and 1414) of spin
susceptibility matrix with (upper) and without (lower) the appearance
of the $\varepsilon$-FS. To produce the $\varepsilon$-FS, the on-site
energy of orbital $d_{xy}$ is adjusted from 0.3 to 0.38, and $\mu$ is
adjusted from 0 to 0.006 to keep the band filling at $n=6$ in the
upper panel, where the divergent susceptibility is $\chi^s_{1414}$ at
$T=0.019$ and $U=1.38$. In the lower panel, $\chi^s_{1414}$ is
suppressed by the annihilation of the $\varepsilon$-FS, and the
divergent channel changes to $\chi^s_{1111}$. With the definition of
pseudo-transition-temperatures $T^*_{d1}$ and $T^*_{d2}$ as
$\chi^s(Q,T^*_{d1(d2)})=10^2\chi^0(Q,T^*_{d1(d2)})$, our numerical
results show that the transition temperature $T^*_{d2}=0.019$ is
higher than $T^*_{d1}=0.012$ in the presence of $\varepsilon$-FS,
which leads to a higher transition temperature of OTDW than that of
SDW, thus the separation between $T_S$ and $T_N$ appears, as is
observed in 1111 compounds. Without $\varepsilon$-FS, we get
$T^*_{d1}=0.017>T^*_{d2}=0.0068$ which indicates same $T_S$ and $T_N$
in our scenario, as is observed in 122 compounds. In this sense, the
essential difference in the structure distortion between 1111 and 122
systems may be attributed to the existence of $\varepsilon$-FS in 1111
parent compounds. Meanwhile, we want to emphasize that depending on
the detailed band structure, there are other possible realizations of
OTDW, such as the inter-orbital nesting between $d_{xz}$ and $d_{yz}$
orbitals. The importance of the $\varepsilon$-FS in splitting $T_S$
and $T_N$ awaits for further experimental verification.

\section{summary}
\label{sum}
We make several remarks as the summary. (1) As has already been
addressed, the OTDW is not a static ordered state of total charge or
spin density, which makes it ``invisible'' to the neutron detections.
(2) The transverse nature of OTDW implies that the ordered orbital
component is a mixture of the two involved orbitals, for instance, a
mixture of $d_{xz}$/$d_{yz}$ and $d_{xy}$ orbitals in the iron-based
superconductors. (3) In our model calculations, the instabilities of
OTDWs occur in the space spanned by $d_{xz}$ (or $d_{yz}$) and
$d_{xy}$ orbitals. However, depending on the band structure, other
possibilities exist. (4) The OTDW might be related with the ``hidden''
order claimed by some experiments~\cite{zabolotnyy:569}. (5) One
intriguing feature stemming from the intinerant multi-orbital model is
that the established SDW should be orbital-polarized (as is shown in
Fig. 3).

\begin{acknowledgements}
The work was supported by the RGC of Hong Kong under Grant No.
HKU7055/09P, the URC fund of HKU, the NSFC (No. 10525415 and No.
10674179), and the 973 Program of China (No. 2006CB921800).
\end{acknowledgements}


\begin{thebibliography}{32}
\bibitem[{hyb()}]{hybri}
\bibinfo{note}{The hybridization parts of Fig. 1 (a2) and (b) are assumed to be
  neglecting small, so we can safely ignore the off-diagonal blocks of the
  susceptibility matrices.}
\end{thebibliography}
\end{document}